\documentclass[graybox]{svmult}

\usepackage{mathptmx}       
\usepackage{helvet}         
\usepackage{courier}        
\usepackage{type1cm}        
\usepackage{makeidx}         
\usepackage{graphicx}        
\usepackage{multicol}        
\usepackage[bottom]{footmisc}

\newcommand{\msun}{M_{\odot}}

\makeindex             

\begin{document}

\title*{Pulsar timing arrays and the challenge of massive black hole binary astrophysics}
\author{A. Sesana$^1$}
\institute{$^1$\ Max-Planck-Institut f\"ur Gravitationsphysik, Albert Einstein Institut, Am M\"ulenber 1, 14476 Golm, Germany, \email{alberto.sesana@aei.mpg.de}}
%
%
\maketitle

\abstract{Pulsar timing arrays (PTAs) are designed to detect gravitational waves (GWs) at nHz frequencies. The expected dominant signal is given by the superposition of all waves emitted by the cosmological population of supermassive black hole (SMBH) binaries. Such superposition creates an incoherent stochastic background, on top of which particularly bright or nearby sources might be individually resolved. In this contribution I describe the properties of the expected GW signal, highlighting its dependence on the overall binary population, the relation between SMBHs and their hosts, and their coupling with the stellar and gaseous environment. I describe the status of current PTA efforts, and prospect of future detection and SMBH binary astrophysics.}

\section{Introduction}
Pulsar timing arrays (PTAs) provide a unique opportunity to obtain the very first low-frequency gravitational wave (GW) detection. The European Pulsar Timing Array (EPTA) \cite{kramer13}, the North American Nanohertz Observatory for Gravitational Waves (NANOGrav) \cite{mclaughlin13}, and the Parkes Pulsar Timing Array (PPTA) \cite{hobbs13}, joining together in the International Pulsar Timing Array (IPTA) \cite{hobbs10,manchester13}, are constantly improving their sensitivity in the frequency range of $\sim10^{-9}-10^{-6}$ Hz, where inspiralling supermassive black hole (SMBH) binaries populating merging galaxies throughout the Universe are expected to generate a strong signal \cite{rr95,jaffe03,wl03,sesana04}. 

Despite the fact that theoretical models of galaxy formation in the standard hierarchical framework predict a large population of SMBH binaries forming during galaxy mergers, to date there is only circumstantial observational evidence of their existence. About 20 SMBH pairs with separations of $\sim10$ pc to $\sim10$ kpc are known as of today (see \cite{dotti12}, for a comprehensive review), among which a new discovered SMBH triple \cite{deane14}. PTAs are sensitive to more compact systems, namely sub-parsec bound Keplerian SMBH binaries. Only few candidates of this class have been identified, based on peculiar broad emission line shifts \cite{tsalmantza11,eracleous11}; however, alternative explanations to the binary hypothesis exist \cite{dotti12}, and unquestionable observational evidence of their binary nature is still missing.

In this contribution I review the properties of the GW signals relevant to PTAs. I describe the status of current PTA efforts, future detection prospects and related astrophysical payouts. If as abundant as predicted, SMBH binaries are expected to form a low frequency background of GWs with a typical strain amplitude $A\sim10^{-15}$ at a frequency $f=1\,$yr$^{-1}$ \cite{sesana08,ravi12,mcwilliams14}, with a considerable uncertainty of $\approx0.5\,$dex{\footnote{In astronomy, the notation dex is commonly used for the log$_{10}$ unit; therefore $0.5\,$dex$=10^{0.5}$.}}. The aforementioned studies indicate that the signal is expected to be dominated by a handful of sources, some of which might be individually resolvable; a situation similar to the foreground generated by WD-WD binaries in the mHz regime relevant to spaced based interferometers like eLISA \cite{gu13}. On the one hand, the unresolved background provides ways to probe the overall population of SMBH binaries in the low redshift universe ($z<1$); on the other hand, electromagnetic counterparts to individually resolvable sources can be searched for with a number of facilities opening new avenues toward a multimessenger based understanding of these fascinating systems and their hosts.


The manuscript is organized as follows. In Section 2, I introduce the general response of a PTA to a passing GW, focusing in Section 3 on the overall stochastic background generated by a population of SMBH binaries. In section 4 I describe the influence of SMBH astrophysics (dynamical coupling with the environment, eccentricity evolution) on the expected signal, and the related issues that a putative PTA detection might answer in the near future. Section 5 is devoted to individually resolved sources and multimessenger astronomy. A summary of the main results is given in Section 6. A concordance $\Lambda$--CDM universe with $\Omega_M=0.27$, $\Omega_\lambda=0.73$ and $h=0.7$ is assumed. Unless otherwise specified, equations are expressed in geometric units where $G=c=1$.

\section{Pulsar timing array response to gravitational waves}
GWs affect the propagation of radio signals from the pulsar to the receiver on Earth \cite{sazhin}, leaving a characteristic fingerprint in the time of arrival of the radio pulses \cite{hd83,j05}. A binary system of masses $M_1>M_2$ in circular orbit at a Keplerian frequency $f_{K}=(1/2\pi)\sqrt{(M_1+M_2)/a^3}$ ($a$ is the semimajor axis of the binary) generates a time dependent GW at a frequency $f=2f_{K}$. In the quadrupolar approximation (sufficient to our scopes) the wave can be described in terms of two independent polarization amplitudes in the form (see, e.g., \cite{sesanavecchio10}):
\begin{eqnarray}
h_+(t) & = A_\mathrm{gw} a(\iota) \cos\Phi(t)\,,\nonumber\\
h_{\times}(t) &= A_\mathrm{gw} b(\iota) \sin\Phi(t)\,,
\end{eqnarray}
where
\begin{equation}
A_\mathrm{gw}(f) = 2 \frac{{\cal M}^{5/3}}{D}\,\left[\pi f(t)\right]^{2/3}
\label{e:Agw}
\end{equation}
is the GW amplitude, ${\cal M}=(M_1M_2)^{3/5}/(M_1+M_2)^{1/5}$ is the chirp mass of the system, $D$ the luminosity distance to the GW source, $\Phi(t)$ is the GW phase $\Phi(t) = 2\pi\int^t f(t') dt'$\,, and $f(t)$ the instantaneous GW frequency {\footnote{The frequency evolution of nHz SMBH binaries over an observing time of $\sim10\,$yr is negligible \cite{sesanavecchio10}, we therefore assume monochromatic non-evolving sources and drop the time dependence in the frequency.}}. The two functions $a(\iota)  = 1 + \cos^2 \iota$ and $b(\iota) = -2 \cos\iota$ depend on the source inclination angle $\iota$. 

When crossing the Earth-pulsar line of sight, such GW induces an 'effective redshift' in the frequency of the received pulses, associated to the oscillatory change in the spacetime background metric, which is proportional to the amplitude $h$ of the wave:
\begin{equation}
z(t,\Omega)  \equiv  \frac{\nu(t) - \nu_0}{\nu_0}=F_+(\Omega)\Delta{h_+}(t)+F_\times(\Omega)\Delta{h_\times}(t).
\label{red}
\end{equation}
Here $F_+(\Omega)$ and $F_\times(\Omega)$ are response functions that depend on the angle defined by the relative pulsar-source orientations in the sky, and the quantities $\Delta{h_{+,\times}}(t)={h_{+,\times}}(t_p)-h_{+,\times}(t_E)$ take into account of the 'double response' of the pulsar-Earth system to the passing GW: what matters is the difference between the metric perturbation at the pulsar and the metric perturbation on Earth. The integral over the observation time of equation (\ref{red}) results in a dephasing of the received pulse: i.e. {\it the GW causes the photons to arrive a little earlier or later than expected, inducing an effective 'residual' in the pulse time of arrival:}
\begin{equation}
r(t) = \int_0^t dt' z(t',\Omega)\,.
\end{equation}
 Being the integral of a sinusoidal wave of amplitude $h$ at a frequency $f$, the residual has a typical magnitude $r\sim h/(2\pi f)$. This relation can be used to conveniently convert strain amplitudes $h$ (which we use in this contribution) into residuals $r$. To get a sense of the magnitude of the effect, the residual can be normalized to fiducial SMBH binary values as
\begin{equation}
r \simeq 30\, \left(\frac{{\cal M}}{10^9\,\msun}\right)^{5/3}\,\left(\frac{D}{100\,\mathrm{Mpc}}\right)^{-1} \times \left(\frac{f}{5\times 10^{-8}\,\mathrm{Hz}}\right)^{-1/3}\,\mathrm{ns},
\label{e:alpha}
\end{equation}
i.e. 100ns or better timing precision is needed in order to make a detection. 

Millisecond pulsars are the most stable natural clocks in the Universe. Nevertheless, timing them to 100ns precision is challenging and several sources of uncontrolled noise might affect the process \cite{verbiest09,shannon10,cordes13}. In particular, red noise processes that cannot be perfectly modeled and subtracted can mimic a low frequency GW signal. It is therefore necessary to monitor an ensemble, or array, of pulsars -- which is the very essence of the concept of pulsar timing array \cite{foster90} --. Only by observing a signal that is consistently cross-correlated among them \cite{hd83} one can be sure of its GW nature as opposed to some intrinsic random noise process.

In observations with PTAs, radio-pulsars are monitored weekly for periods of years. The relevant frequency band is therefore between $1/T$ -- where $T$ is the total observation time -- and the Nyquist frequency $1/(2\Delta t)$ -- where $\Delta t$ is the time between two adjacent observations --, corresponding to $3\times 10^{-9}$ Hz - few$\times10^{-7}$ Hz. The frequency resolution bin is $1/T$. 

\begin{figure}
\centering
\includegraphics[width=4.5in]{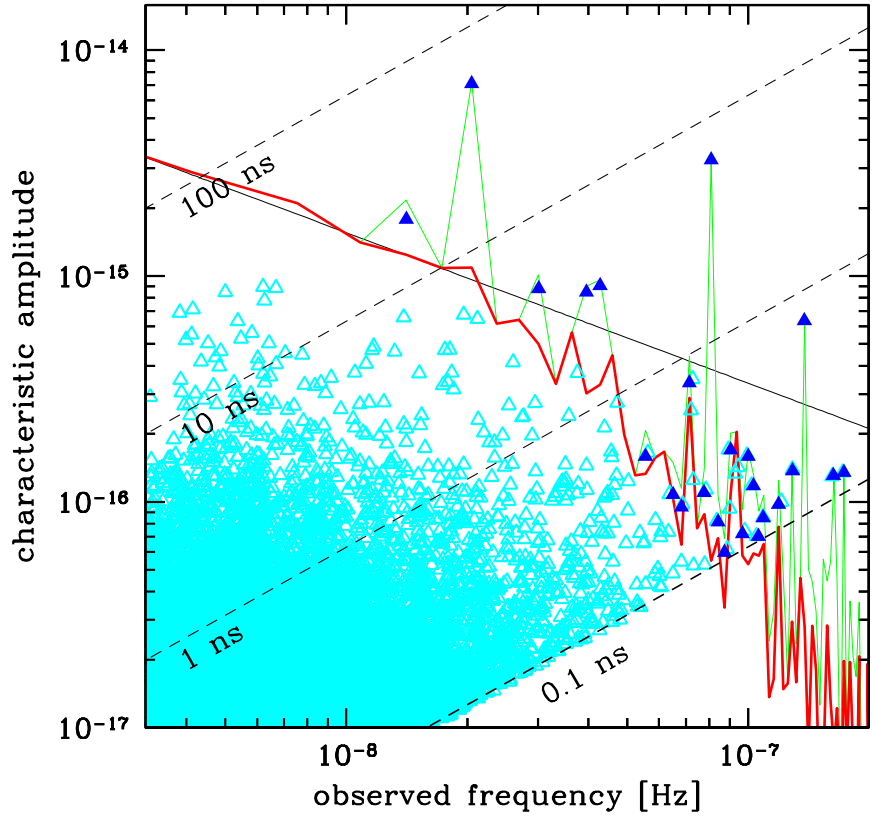}

\caption{Illustrative realization of the overall GW signal in the frequency domain; characteristic amplitude $h_c$ vs frequency. Each cyan triangle corresponds to the contribution of an individual SMBH binary; among these, blue triangles identify bright, resolvable sources. The overall GW signal is given by the jagged green line, whereas the red thick line represents the unresolved signal after subtraction of the brightest sources. The solid black line represents the theoretical $h_c\propto f^{-2/3}$ behavior, and dashed lines mark characteristic residual levels according to the conversion $r=h/(2\pi f)$, as labeled in figure.}
\label{fig1}
\end{figure}

Given a cosmological population of SMBH binaries, the overall GW signal reaching the Earth is an incoherent superposition of monochromatic waves $\sum_i h_i$. Depending on the details of the population, such superposition can behave as an effective 'stochastic background', and can be dominated by few bright resolvable sources. This concept is represented in figure \ref{fig1}. For the selected realization of the SMBH population, the signal looks like a stochastic background at $f<10^{-8}$Hz; it is however dominated by bright resolvable sources at higher frequencies. We will first treat the overall signal as an effective background composed of unresolvable sources, we will then focus on the properties of individual sources.

\section{The stochastic GW background}
We now consider a cosmological population of inspiralling SMBH binaries. Without making any restrictive assumption about the physical mechanism driving the binary semimajor axis -- $a$ -- and eccentricity -- $e$ -- evolution, we can write the characteristic amplitude $h_c$ of the GW signal generated by the overall population as \cite{sesana13b}:
\begin{eqnarray}
h_c^2(f) = &\int_0^{\infty}dz\int_0^{\infty}dM_1\int_0^{1}dq \frac{d^4N}{dzdM_1dqdt_r}\frac{dt_r}{d{\rm ln}f_{{\rm K},r}}\times\nonumber\\
& h^2(f_{{\rm K},r})\sum_{n=1}^{\infty}\frac{g[n,e(f_{{\rm K},r})]}{(n/2)^2}\,\delta\left[f-\frac{nf_{{\rm K},r}}{1+z}\right].
\label{hch2}
\end{eqnarray}
Here $h(f_{{\rm K},r})$ is the strain emitted by a circular binary at a Keplerian rest frame frequency $f_{{\rm K},r}$ {\footnote{For a source at redshift $z$, the general relation between restframe and observed frequency is $f_r=f(1+z)$.}}, averaged over source orientations
\begin{equation} 
h(f_{{\rm K},r})=\sqrt{\frac{32}{5}}\frac{{\cal M}^{5/3}}{D}(2\pi f_{{\rm K},r})^{2/3}.
\label{hrms}
\end{equation}
The function $g(n,e) $ \cite{pm63} accounts for the fact that the binary radiates GWs in the whole spectrum of harmonics $f_{r,n}=nf_{{\rm K},r}\,\,\,(n=1, 2, ...)$, and is given by, e.g., equations (5)-(7) in \cite{amaro10}. The $\delta$ function, ensures that each harmonic $n$ contributes to the signal at an observed frequency $f=nf_{{\rm K},r}/(1+z)$, where the factor $1+z$ is given by the cosmological redshift. $d^4N/(dzdM_1dqdt_r)$ is the differential cosmological coalescence rate (number of coalescences per year) of SMBH binaries per unit redshift $z$, primary mass $M_1$, and mass ratio $q=M_2/M_1<1$. $dt_r/d{\rm ln}f_{{\rm K},r}$ is the time spent by the binary at each logarithmic frequency interval. These two latter terms, taken together, simply give the instantaneous population of comoving systems orbiting at a given logarithmic Keplerian frequency interval per unit redshift, mass and mass ratio. In the case of circular GW driven binaries, $g(n,e)=\delta_{n2}$, $dt/d{\rm ln}f$ is given by the standard quadrupole formula (see equation (\ref{e:fdot}) below), and equation (\ref{hch2}) reduces to the usual form \cite{phinney01} 
\begin{equation}
h_c^2(f) =\frac{4f^{-4/3}}{3\pi^{1/3}}\int \int dzd{\cal M} \, \frac{d^2n}{dzd{\cal M}}{1\over{(1+z)^{1/3}}}{\cal M}^{5/3},
\label{hcirc}
\end{equation}
where we have introduced the differential merger remnant density (i.e. number of mergers remnants per co moving volume) $d^2n/(dzd{\cal M})$ (see \cite{phinney01,sesana13} for details). In this case, $h_c\propto f^{-2/3}$; it is therefore customary to write the characteristic amplitude in the form $h_c=A(f/{\rm yr}^{-1})^{-2/3}$, where $A$ is the amplitude of the signal at the reference frequency $f=1 {\rm yr}^{-1}$. Observational limits on the GW background are usually given in terms of $A$.

\section{The GW background as a tool for SMBH binary astrophysics}
Equation (\ref{hch2}), together with a prescription for the eccentricity distribution of the emitting SMBH binaries as a function of frequency, namely $e(M_1,q,f_{{\rm K,r}})$, provides the most general description of the GW background generated by a population of SMBH binaries. The signal depends on three distinctive terms:
\begin{enumerate}
\item the cosmological SMBH binary coalescence rate , $d^4N/(dzdM_1dqdt_r)$;
\item the frequency evolution of each binary, $dt_r/d{\rm ln}f_{{\rm K},r}$;
\item the eccentricity evolution of the systems, which determines the emitted spectrum at any given binary Keplerian frequency.
\end{enumerate}
In the following section, we will examine the impact of the items listed above on the GW signal, highlighting the enormous potential of PTA observations to enhance our understanding of the global population and the dynamical evolution of SMBH binaries in the Universe.

\subsection{Spectral amplitude: the cosmic SMBH binary merger rate}
We first consider GW-driven, circular SMBH binaries only. It is clear by looking at equation (\ref{hcirc}) that the GW strain amplitude is proportional to the square root of the cosmic coalescence rate of SMBH binaries, and it is sensitive to their mass distributions. Therefore, the coalescence rate sets the {\it normalization} of the detectable signal. As discussed in \cite{sesana13} this, in practice, depends on four ingredients: (i) the galaxy merger rate; (ii) the relation between SMBHs and their hosts, (iii) the efficiency of SMBH coalescence following galaxy mergers and (iv) when and how accretion is triggered during a merger event. Our limited knowledge of each of these ingredients is reflected in the uncertainty range of the predicted GW signal amplitude. 

The expected level of the signal under the circular GW-driven approximation has been investigated by several authors \cite{sesana08,ravi12,sesana13,ravi14b,mcwilliams14}, producing largely consistent results that cluster around the range $5\times10^{-16}<A<2\times10^{-15}$. In particular Sesana and collaborators demonstrated that different methodologies that can be used in the computation of the signal yield consistent results \cite{sesana08,sesana09,sesana13}. Ravi and collaborators corroborated those findings with independent investigations based on the Millennium Run \cite{springel05}, and by constructing an alternative observation-based model \cite{ravi14b}. \cite{mcwilliams14} finds a somewhat larger signal, of the order $A\sim4\times10^{15}$. This may be ascribed to a set of assumption, including a non evolving SMBH density with cosmic time and a merger driven only evolution of the SMBH mass function, that tend to favor more and more massive SMBH binary mergers.

\begin{figure}
\centering
\includegraphics[width=4.5in]{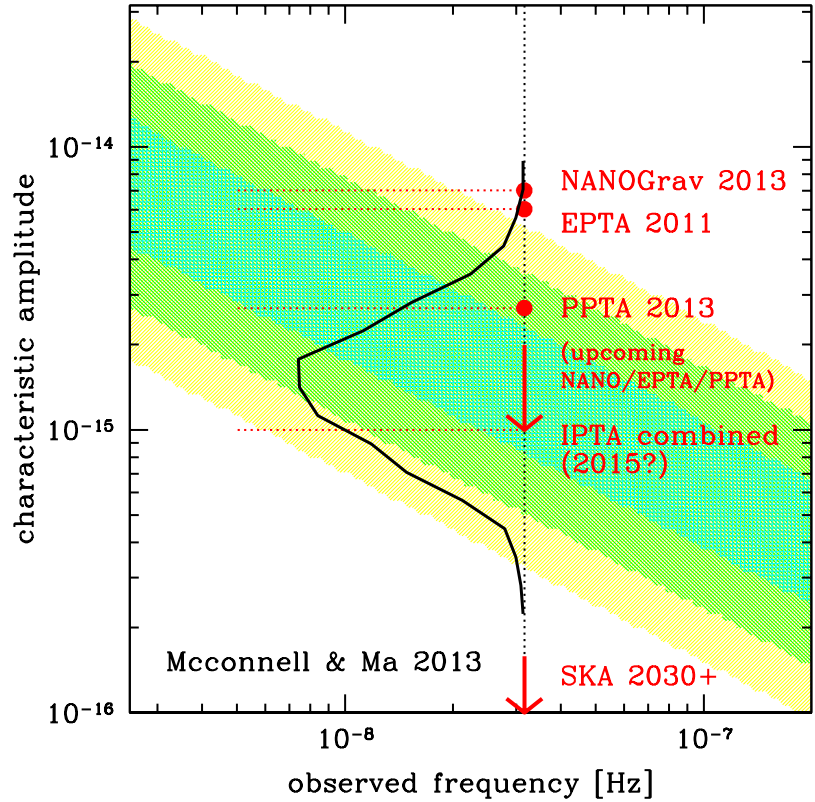}
\caption{Characteristic amplitude of the GW signal. Shaded areas represent the  $68\%$, $95\%$ and $99.7\%$ confidence levels given by a representative set of observation-based models from \protect\cite{sesana13}, and the overlied solid black line traces the probability distribution function of $A$ (i.e. $h_c$ at $f=1$yr$^{-1}$). The red dots mark the progression of PTA limits from \protect\cite{vanh11,demorest13,shannon13}. The two red arrows anticipate the expected sensitivity of upcoming combined IPTA datasets, and of the full SKA \protect\cite{dewdney09}.}
\label{fig2}
\end{figure}

Figure \ref{fig2} compares the predicted GW amplitude to current observations.  Here we plot $68\%$, $95\%$ and $99.7\%$ confidence level of the expected characteristic strain, extracted from a large compilation of models featuring different prescriptions for the SMBH binary population, as described in \cite{sesana13}. In particular, we consider here the SMBH-bulge scaling relations derived in \cite{mcconnell13}, which are representative of a recent upward revision of those relations \cite{kormendy13}. It is worth mentioning how the most stringent upper limit published to date ($A>2.4\times10^{15}$ at 95\% confidence, \cite{shannon13}) is skimming the predicted $68\%$ confidence region. Upcoming limits from all the PTAs might soon push the numbers down by few dex to $A\approx2\times10^{15}$, and the combined IPTA dataset might be sensitive to $A$ as small as $10^{-15}$, implying a good chance of detection in the next few years.  

The first, obvious payout of a PTA detection is the {\it direct} confirmation of the existence of a vast population of sub-pc (to be precise, sub-$0.01$pc) SMBH binaries. From this we learn that (i) binaries efficiently pair on pc scales following galaxy mergers and (ii) stellar and/or gas dynamics is effective in removing energy and angular momentum from the binary, overcoming the 'last parsec problem'. It can still be the case that some level of stalling may occur, delaying the efficient energy loss due to GW emission. Nevertheless, a direct PTA detection will confirm that SMBH binaries eventually coalesce within an Hubble time. The exact level of the signal then depends on the combined ingredients outlined above, and those cannot be constrained all together by measuring one single number. It is even more dangerous to try to draw conclusions about SMBH coalescence rates based on PTA upper limits. In the following section we will discuss how environmental coupling can result in a loss of low frequency GW power; a non-detection at a given level might be either due to a particularly low coalescence rate, or to a particularly efficient coupling to the environment.

\subsection{Spectral shape: the SMBH binary-environment coupling}
SMBH binaries do not evolve in isolation, and although the circular GW-driven approximation has been widely employed to make predictions, GW emission alone cannot bring a SMBH binary to coalescence. This is clear by looking at the 'residence time' $dt_r/d{\rm ln}f_r$, i.e., the time a binary spend at each logarithmic frequency bin. In the circular GW-driven approximation, this is 
\begin{equation}
\frac{dt_r}{d{\rm ln}f_r} = \frac{5}{64\pi^{8/3}} {\cal M}^{-5/3}f_r^{-8/3}.
\label{e:fdot}
\end{equation}
-- note that this, combined to equation (\ref{hrms}), yields $h_c\propto M_1^{5/6}q^{1/2}f^{-2/3}$, i.e. the standard $f^{-2/3}$ power law given by equation (\ref{hcirc}) --. We note the steep frequency dependence of equation (\ref{e:fdot}): as a matter of fact, typical binary residence times become longer than the Hubble time for $f<10^{-10}$Hz (roughly semimajor axis $a>0.1$pc for the $M\approx10^{9}\msun$ systems relevant here). It is therefore clear that some mechanism other then GW emission must bring the binary to subparsec separations. 

Forming after galaxy mergers, SMBH binaries sit at the center of the stellar bulge of the remnant, and they are possibly surrounded by massive gas inflows triggered by dynamical instabilities related to the strong variations of the gravitational potential during the merger episode. Accordingly, two major routes for the SMBH binary dynamical evolution have been explored in the literature: (i) gas driven binaries, and (ii) stellar driven binaries. A detailed description of both scenarios is beyond the scope of this contribution; here we consider simple evolutionary routes and assess their impact on the GW signal. 

\subsubsection{Change in the SMBH binary shrinking rate}
Let restrict ourselves to circular binaries first. A background of stars scattering off the binary drives its semimajor axis evolution according to the equation \cite{quinlan96}
\begin{equation}
\frac{da}{dt} = \frac{a^2G\rho}{\sigma}H,
\label{adotstar}
\end{equation}
where $\rho$ is the density of the background stars, $\sigma$ is the stellar velocity dispersion and $H$ is a numerical coefficient of order 15 \cite{quinlan96}. Although equation (\ref{adotstar}) has been originally derived for static uniform backgrounds, recent numerical simulations \cite{khan11,preto11} suggest that it can be applied to realistic systems providing  that the density $\rho$ is evaluated at the binary influence radius ($r_i\approx GM/\sigma^2$). If we consider, for simplicity, an isothermal sphere (i.e. $\rho\propto r^{-2}$), we substitute $\rho_i$ in equation (\ref{adotstar}), and we assume $M_{BH}\propto\sigma^5$, we get $dt/d{\rm ln}f\propto f^{2/3}M_1^{2/3}$, which yields to a contribution of the single binary to the GW background of the form $h_c\propto M_1^2qf$. 

In the case of circumbinary disks, things are even more subtle, and the detailed evolution of the system depends on the complicated and uncertain dissipative physics of the disk itself. The simple case of a coplanar prograde disk, with a central cavity maintained by the torque exerted by the binary onto the disk admits a selfconsistent, non stationary solution that can be approximated as \cite{ipp99,haiman09}
\begin{equation}
\frac{da}{dt} = \frac{2\dot{M}}{\mu}(aa_0)^{1/2}.
\label{adotgas}
\end{equation}
Here, $\dot{M}$ is the mass accretion rate at the outer edge of the disk, $a_0$ is the semimajor axis at which the mass of the unperturbed disk equals the mass of the secondary black hole, and $\mu$ is the reduced mass of the binary. Considering a standard geometrically thin, optically thick disk model \cite{ss73}, one finds $dt/d{\rm ln}f\propto f^{-1/3}M_1^{1/6}$, which yield to a contribution of the single binary  to the GW background of the form $h_c\propto M_1^{7/4}q^{3/2}f^{1/2}$. 

When compared to the GW driven case, $(da/dt)_{gw}\propto a^{-3}$, equations (\ref{adotstar}) and (\ref{adotgas}) provide the transition frequency between the external environment driven and the GW driven regimes:
\begin{equation}
f_{{\rm star/gw}}\approx 5\times10^{-9}M_8^{-7/10}q^{-3/10}\,{\rm Hz},
\label{decoupstar}
\end{equation}
\begin{equation}
f_{{\rm gas/gw}}\approx 5\times10^{-9}M_8^{-37/49}q^{-69/98}\,{\rm Hz},
\label{decoupgas}
\end{equation}
where $M_8=M/10^8\msun$. We therefore see that if the signal is dominated by $10^{9}\msun$ SMBH binaries, then the transition frequency is located around $10^{-9}$Hz. 

\subsubsection{Eccentricity evolution}
It is well known that GW emission efficiently circularizes binaries, however things can be dramatically different in the star and gas dominated stages. If binaries get very eccentric in those phases, they can retain substantial eccentricity even during the GW dominated inspiral relevant to PTA observations, beyond the decoupling frequencies given by equations (\ref{decoupstar}) and (\ref{decoupgas}). The eccentricity evolution in stellar environments has been tackled by several authors by means of full N-body simulations and semianalytic models. A consistent picture emerged according to which equal mass, circular binaries tend to stay circular or experience a mild eccentricity increase, while binaries that form already eccentric, or with $q\ll 1$ (regardless of their initial eccentricity) tend to grow more eccentric \cite{mms07,mat07,preto11,quinlan96,sesana06,sesana10}. An important parameter here seems to be the eccentricity of the binary at the moment of formation
$e_0$, which is often found to be larger than 0.6 in numerical studies \cite{preto11}. Large $e_0$ implies that systems emitting in the nHz regime can be highly eccentric, causing a significant suppression of the GW signal, as we will see in the next section. In the circumbinary disk scenario, excitation of eccentricity has been seen in several simulations \cite{armitage05,cuadra09}. In particular, the existence of a limiting eccentricity $e_{\rm crit} \approx 0.6 - 0.8$ has been reported in \cite{roedig11} through a suite of high resolution smoothed particle hydrodynamics simulations, in the case of massive selfgravitating disks. Therefore, also in gaseous rich environments, eccentric binaries might be the norm -- even though the extreme eccentricities ($e>0.9$) that might be reached in the stellar driven case are unlikely --.

\subsubsection{Impact on the GW signal}
\begin{figure}
\centering
\resizebox{\hsize}{!}{\includegraphics[clip=true]{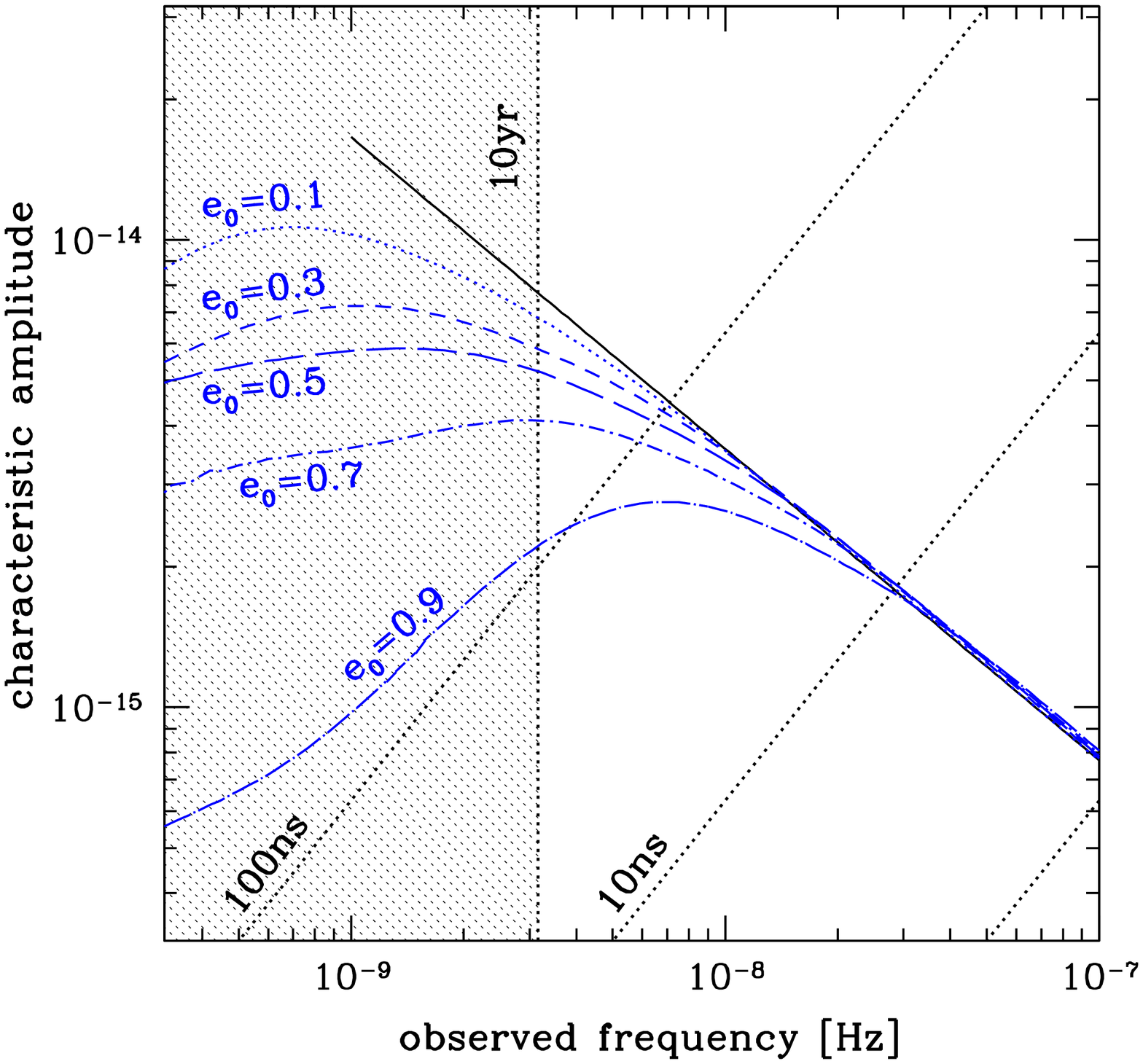}
\includegraphics[clip=true]{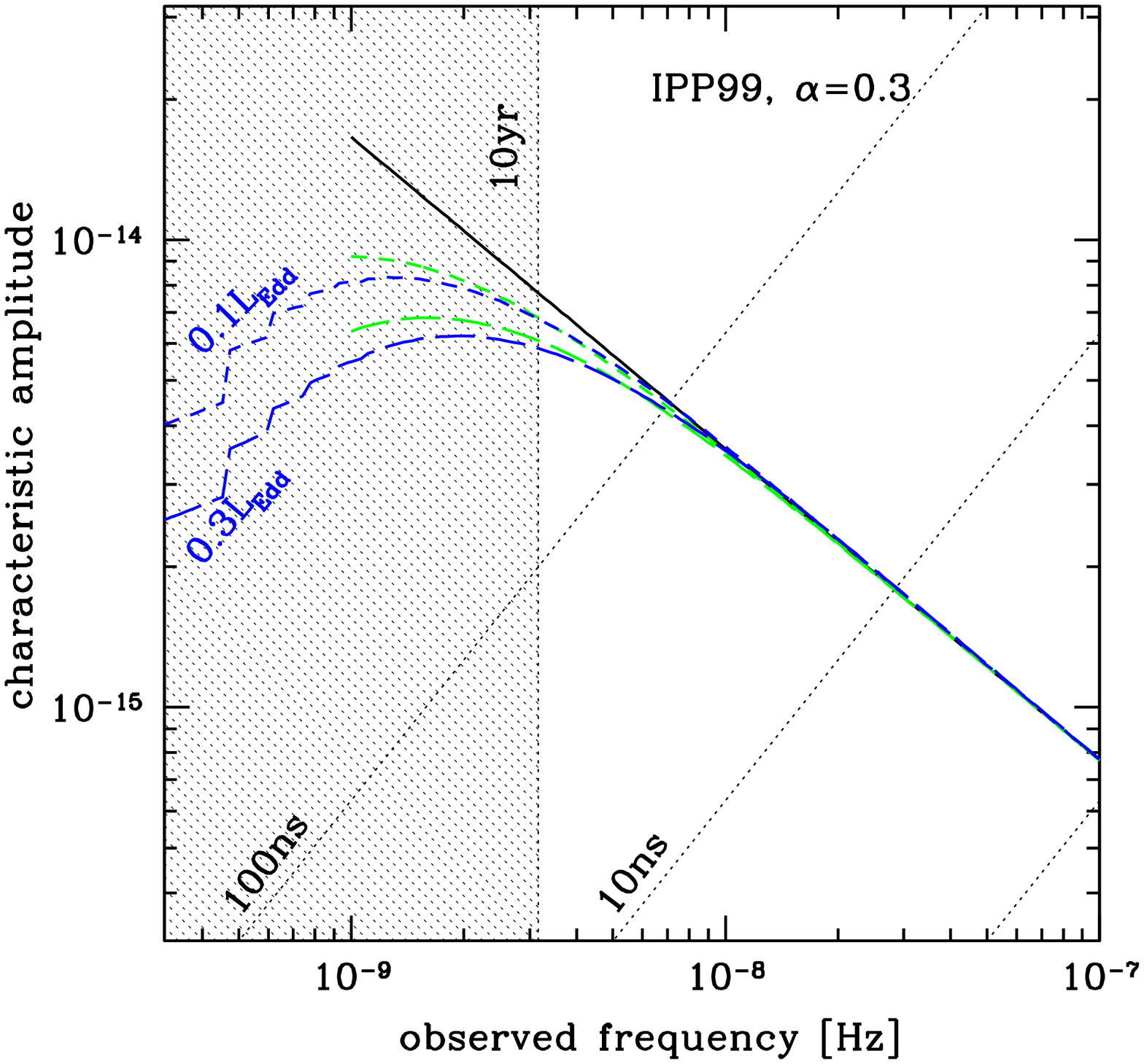}}
\caption{Effect of the environmental coupling on the overall GW spectrum. {\it Left panel}: SMBH binary evolution in a stellar background. Blue lines represent the signal generated by a population of binaries evolving in a dense stellar background, according to the model presented in \protect\cite{sesana10}; each line is for a specific fiducial eccentricity at binary formation -- $e_0$ --, as labeled in figure. {\it Left panel}: SMBH binary evolution in circumbinary gaseous disks. Here the two green curves assume circular binaries, whereas the blue curves allow self-consistent eccentricity evolution (resulting in a slightly lower signal). The upper and lower pair of curves are for two different Eddington ratios (see text), as labeled in figure. In both panels, the shaded area cuts off the region at $f<3\times10^{-9}$, to highlight the modified signal in a frequency range that is relevant to a $\sim 10$yr observation time. As in figure \ref{fig1}, the solid black lines represent the theoretical $h_c\propto f^{-2/3}$ behavior, and dashed lines mark residual levels according to the conversion $r=h/(2\pi f)$, as labeled in figure.}  
\label{fig3}
\end{figure}
The main effect of the binary-environment coupling is to {\it suppress} the low frequency signal \cite{sesana04,enoki07,kocsis11,ravi14a}. On the one hand, the energy of the SMBH binary is transferred to the environment instead of going into GWs, the binary evolution is faster, and consequently there are less systems emitting at each frequency. On the other hand, if eccentricity grows significantly, GW power is moved from the second to higher harmonics, through the function $g(n,e)$ \cite{pm63}. However, since the energy carried by the wave is proportional to $f^2h^2$, shifting the emission to higher harmonics effectively {\it removes} power at low frequencies, without a significant enhancement (just marginal) of $h$ at higher frequencies \cite{enoki07}. Therefore, generally speaking, highly eccentric binaries might pose a threat to PTA GW detection \cite{sesana13,ravi14a}. 

This is clearly shown in figure \ref{fig3}, for selected systems evolving in stellar (left panel) and gaseous (right panel) environments. In both cases, the transition frequency is around $10^{-9}$ Hz (corresponding to 30 yr timescale), and has only a marginal influence on detectability (assuming 10 yr of observation). However, if the PTA baseline is further extended in time, departures from the standard $f^{-2/3}$ power law become significant. The situation may get problematic when binaries grow very eccentric. This is particularly true in the stellar driven case, where, if binaries form already quite eccentric (say $e_0>0.7$), eccentricity can easily grow to $e>0.95$. Consequently, a large amount of power at low frequency is removed, as shown by the $e_0=0.9$ line in the left panel of figure \ref{fig3}. Note that in this case, the suppression of the signal at low frequency is so severe that even 10yr observations with a nominal array sensitivity of 100ns would not be sufficient for a confident detection. This problem is generally less severe in gaseous driven systems, where the existence of a limiting eccentricity (at least in the prograde case) prevents significant loss of power at low frequencies (as shown by the blue curves in the right panel of figure \ref{fig3}). 

We must caution, however, that these results are based on toy models that probably tend to overestimate the impact of the binary coupling to the environment. The isothermal sphere implies central stellar densities higher than those typically seen in large ellipticals (see, e.g.m \cite{terzic05}), which accelerates the pace at which the binary shrinks according to equation (\ref{adotstar}). Also in the gas driven case, the evolution described by equation (\ref{adotgas}) is function of the accretion rate. Here we considered binaries accreting at $0.1-0.3$ of the Eddington rate \footnote{We recall that the Eddington luminosity is the maximum admitted luminosity for which the radiation pressure exerted by the photons emitted in the accretion process is smaller than the gravitational binding energy of the accreting material. If the contrary is true, radiation pressure blows away the reservoir of gas, suppressing the accretion process. For standard radiatively efficient accretion flows \cite{ss73}, the Eddington luminosity corresponds to an Eddington accretion rate of $\dot{M}_{\rm Edd}\approx2.5 M_8\msun$yr$^{-1}$, where $M_8$ is the MBH mass normalized to 10$^8\msun$. Accretion rates are customarily given as fractions of the Eddington rate.}. However, observations of active SMBHs at low redshift indicate that the most massive systems (most relevant to PTA) are generally severely sub-Eddington \cite{kauffmann09}, which implies a weaker binary-environment coupling. It is, however, clear that the determination of the GW background spectral slope carries a lot of information about the dynamics of SMBH binaries. A well defined turnover frequency around $10^{-9}$ Hz will be the distinctive signature that strong coupling with the environment is the norm, whereas a plateau might be indicative of a population of highly eccentric systems. PTA detection will therefore provide important information about the dynamics of individual SMBH binaries, not only about the statistics of their collective population.

\section{Resolvable sources}
As shown in figure \ref{fig1}, a handful of sources might be bright enough to be individually resolved, i.e., they stand out of the total signal level produced by the combination of all other systems. This can be seen by computing the signal-to-noise ratio (SNR) of a bright source with respect to the instrumental noise plus the unresolved GW background. Assuming a monochromatic source, the SNR in a single pulsar $\alpha$ can be approximated as \cite{sesanavecchio10}
\begin{equation}
\rho_\alpha^2=(r_\alpha | r_\alpha)\approx\frac{2}{S_\alpha}\int_0^{T} r^2_\alpha(t) dt,
\end{equation}
where the inner product $(*|*)$ is defined by the last approximate equality and $S_0$ in the noise spectral density evaluated at the frequency of the source, including both the pulsar noise and the unresolved GW background contribution. For an array of $M$ pulsars, the single contribution of each pulsar can be coherently added in quadrature to get $\rho^2=\sum_{\alpha = 1}^M \rho_\alpha^2$. 

The statistics of resolvable sources has only been investigated by \cite{sesana09}. They found that an handful of sources might be resolvable on top of the stochastic background, which seems to be corroborated by recent work from Ravi and collaborators \cite{ravi14b}. This is an active area of research at the moment; several single source data analysis pipelines are under construction, and a better understanding of the typical properties of resolvable SMBH binaries will help setting the requirements for detection algorithm development.

\begin{figure}
\centering
\includegraphics[width=4.5in]{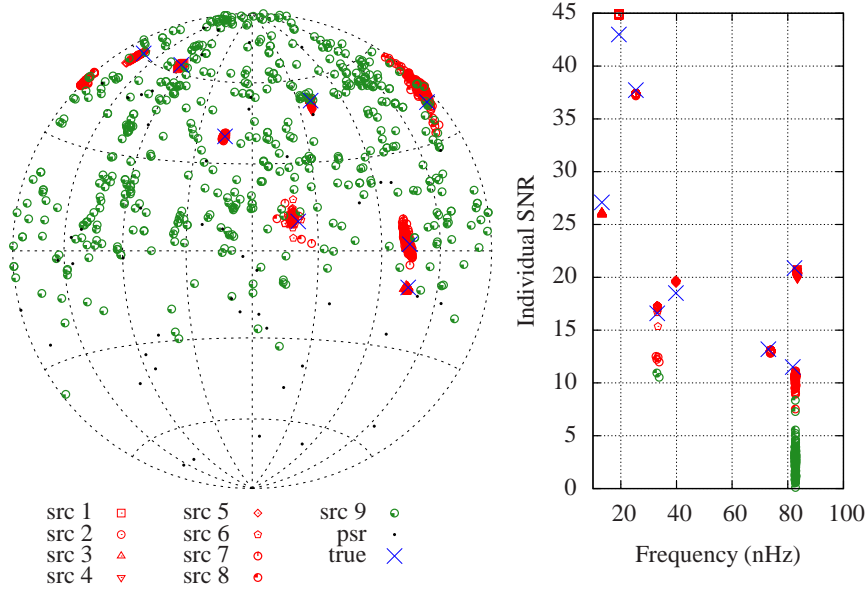}
\caption{Performance of the multi-search genetic algorithm in finding resolvable sources. The figure shows the source localization in the sky (sky map, left panel) and in the frequency-SNR space (right panel). Blue crosses ($\times$) correspond to injected signals and black dots to the position of the millisecond pulsars forming the array. Red marks represent sources 1-to-8 found by all the organisms with SNR$^2_{\rm tot} > 99 \%$SNR$^2_{\rm best}$, while green circles represent a putative 9th source. Note that this latter one does not have a defined position, and typically has SNR$<5$ (from \protect\cite{petiteau13}).}
\label{fig4}
\end{figure}

\subsection{Science with resolvable PTA sources}
A pioneering investigation of the detectability of resolvable PTA sources was performed by \cite{sesanavecchio10} assuming circular, non spinning monochromatic systems, and considering only the coherent 'Earth term' in the analysis. In this case the waveform is function of 7 parameters only: the amplitude $R${\footnote{For monochromatic signals the two masses and the luminosity distance degenerate into a single amplitude parameter.}}, sky location $\theta,\phi$, polarization $\psi$, inclination $\iota$, frequency $f$ and phase $\Phi_0$, defining the parameter vector $\vec{\lambda} = \{R,\theta,\phi,\psi,\iota,f,\Phi_0\}$. \cite{sesanavecchio10} finds that for SNR$=10$, the source amplitude is determined to a $20\%$ accuracy, whereas $\phi,\psi,\iota$ are only determined within a fraction of a radian and sky location within few tens to few deg$^2$ (depending on the number of pulsars in the array). These results were confirmed and extended by \cite{babak12,petiteau13}. In particular \cite{petiteau13} demonstrated that, providing there are enough pulsars in the array, multiple individual sources can be separated. This is shown in figure \ref{fig4}, which is taken from the original paper. Using a multi-search genetic algorithm, the authors succeeded to correctly identify all the eight sources blindly injected in the data, and to infer that there were no other sources with SNR$>5$. Experiments like this were conducted on simulated data including Gaussian noise only, and in a relatively high SNR regime. The effective capabilities of PTAs to identify and separate individual sources in more realistic situations is subject of active work in the field \cite{ellis12,taylor14,ellis14}. 

For bright enough sources (SNR$\approx10$) sky location within few tens to few deg$^2$ is possible (see also \cite{ellis12}), and even sub deg$^2$ determination, under some specific conditions \cite{lee11}. Even though this is a large chunk of the sky, resolvable systems are extremely massive and at relatively low redshift ($z<0.5$), making them suitable targets for electromagnetic follow-up. The possibility of performing multimessenger astronomy with PTA sources has been firstly investigated in \cite{sesana12,tanaka12} (see \cite{burke13,tanakahaiman13} for a comprehensive review). They identified a number of possible signatures, including: periodicity related to the binary orbital period; peculiar emission spectra due to the presence of a central cavity; peculiar K$\alpha$ line profiles; recurring flaring activity (see also \cite{tanaka13}). Given the relatively poor sky localization, suitable hosts must be selected among galaxies falling in a large error box. In general, bright ellipticals in galaxy cluster centers experience a large merger activity at low redshift, and might be the best candidates for periodic monitoring and dedicated pointings \cite{rosado14,simon14}.   

\section{Conclusions}
PTA increasing sensitivities promise the detection of the predicted GW signal produced by a cosmological population of SMBH binaries in the near future. Beyond the obvious excitement of a direct GW observation, detection will carry an enormous wealth of information about these fascinating astrophysical systems, providing a {\it direct unquestionable} evidence of their existence and an empirical prove that the 'final parsec problem' is solved by nature. The determination of the overall signal amplitude and spectral slope will provide important information about the global properties of the SMBH binary population, giving, for example, insights about the relation between SMBH binaries and their hosts and will inform us about the dynamics of SMBH binaries and their stellar and/or gaseous environment, possibly constraining the efficiency of their mutual interaction. Identification and sky localization of individual sources, will add further excitement to the picture, making multimessenger studies of SMBH binaries and their hosts possible. Although the current focus is inevitably on the first detection, we should never forget that pulsar timing arrays will also be groundbreaking astrophysical probes.

\section*{Acknowledgments}
A.S. acknowledges the DFG grant SFB/TR 7 Gravitational Wave Astronomy and by DLR (Deutsches Zentrum fur Luft- und Raumfahrt), and wishes to thank Carlos F. Sopuerta and all the organizers of the 2014 Sant Cugat Forum on Astrophysics (Gravitational Wave Astrophysics).



\bibliographystyle{unsrt}
\bibliography{references}

\end{document}